\documentclass[a4paper]{article}

\usepackage{INTERSPEECH2021}
\usepackage{amsmath}%
\usepackage{amsfonts}%
\usepackage{amssymb}%
\usepackage{graphicx}
\usepackage{url}
\usepackage{booktabs}
\usepackage{multirow}
\usepackage{placeins}
\usepackage{comment}
\usepackage{todonotes}
\usepackage{url}
\usepackage{subcaption}

\usepackage{bm}
\usepackage{textcomp}

\usepackage{math/math_defs}

\title{Representation Learning to Classify and Detect Adversarial Attacks against \\Speaker and Speech Recognition Systems}
\name{Jes\'us Villalba$^{1,2}$\thanks{This research has been supported by DARPA RED under contract HR00112090132.}, Sonal Joshi$^1$, Piotr \.Zelasko$^{1,2}$, Najim Dehak$^{1,2}$}
\address{
$^1$Center for Language and Speech Processing,
  $^2$Human Language Technology Center of Excellence, \\
    Johns Hopkins University, Baltimore, MD}
\email{\{jvillal7,sjoshi12,pzelask2,ndehak3\}@jhu.edu}

\begin{document}

\maketitle
\begin{abstract}
  Adversarial attacks have become a major threat for machine learning applications. There 
  is a growing interest in studying these attacks in the audio domain, e.g, speech and speaker 
  recognition; and find defenses against them. In this work, we focus on using representation 
  learning to classify/detect attacks w.r.t. the attack algorithm, threat model or
  signal-to-adversarial-noise ratio. We found that common attacks in the literature can
  be classified with accuracies as high as 90\%. Also, representations trained to 
  classify attacks against speaker identification can be used also to classify attacks
  against speaker verification and speech recognition. We also tested an attack verification task,
  where we need to decide whether two speech utterances contain the same attack. We observed 
  that our models did not generalize well to attack algorithms not included in the
  attack representation model training.
  Motivated by this, we evaluated an unknown attack detection task.
 We were able to detect unknown attacks with equal error rates of about 19\%, which is promising. 
\end{abstract}

\vspace{-2mm}
\section{Introduction}

Adversarial attacks~\cite{szegedy-iclr14} have become a major threat for machine learning systems including those based on audio such as 
speaker and speech recognition. These attacks consist of adding an imperceptible perturbation to the speech waveform, which
is optimized to change the output of the systems. 

Though adversarial attacks research started in the image domain~\cite{Goodfellow2015,carlini-16,Kurakin2017},
it is recently gaining attention in the audio domain. There are already works that study the 
effectiveness of adversarial attacks against 
speech recognition~\cite{Cisse2017,iter-17,carlini-18,Neekhara2019,Schonherr2019,Qin2019}, 
 and speaker recognition~\cite{kreuk-icassp18,gong-18,Xie2020,Li2020a,Villalba2020a}. Furthermore, 
 some recent works proposed pre-processing defenses~\cite{joshi2021adversarial,zelasko2021adversarial}, intended to remove the adversarial noise from the audio; or using adversarial training to robustify speaker identification networks~\cite{JATI2021101199}.
 
In this work, we investigate how to detect and classify adversarial attacks against audio systems. 
This research has been done in the context of the DARPA RED (Reverse Engineering of Deceptions)\footnote{\tiny\url{https://www.darpa.mil/program/reverse-engineering-of-deceptions}} program. 
The goal of the program is to produce algorithms for automatically identifying the toolchains behind attacks. 
Obtaining such information could help to identify the attackers, their intentions, and decide which defenses might be most effective against those attacks.
Encouraged by the success of
deep learned representations like x-vectors in fields like speaker verification~\cite{snyder-icassp18,Villalba2020}, emotion recognition~\cite{Pappagari2020}, or
speech pathology detection~\cite{Moro-Velazquez2020}; we decided to apply the same class of network architectures to compute embeddings, which can help us decide whether a test recording has been attacked and extract information about the attack. We will denote these embeddings as \emph{attack signatures} to follow
DARPA's nomenclature.

We experimented on three tasks: attack classification, verification, and unknown attack detection. Attack classification
is the task of deciding whether a test utterance is benign or whether it belongs to a set of known attacks. This classification
can be done w.r.t. either the attack algorithm, the threat model, or the perturbation size --e.g., Signal-to-Noise Ratio (SNR). Similarly to speaker verification, the attack verification task is deciding whether two utterances have been attacked in the same way. 
This can be useful when we have a few samples from an attacker using an unknown deception toolchain, and we want to detect new
attacks from this attacker. Finally, the unknown attack detection task is deciding whether an utterance contains an unknown attack, meaning attack which is not included in our training set. This can be useful to detect new attacks so that later we can add them to our database of attacks.

\vspace{-2mm}
\section{Adversarial Attacks}

\subsection{Threat models}
\vspace{-1mm}
Suppose $\xvec\in\real^T$ is a benign audio waveform of length $T$, also called clean, or bonafide. Let $y^{\mathrm{benign}}$ be its true labels. 
An attacker could craft an adversarial example $\xvec'=\xvec+\deltavec$ by adding an imperceptible perturbation $\deltavec$ to the speech waveform. The adversarial perturbation $\deltavec$ is optimized to alter the decision of speech or speaker recognition systems. 
To enforce imperceptibility of the perturbation, some distance metric is minimized or bounded $D(\xvec,\xvec')<\varepsilon$. 
Typically, this is the $L_p$ norm of the perturbation, $D(\xvec,\xvec')=\left\|\deltavec\right\|_p$. The choosing of this metric is usually 
known in the literature as the \emph{threat model} of the attack. In this work, we consider $L_0, L_1, L_2$ and $L_\infty$ threat models. 

\vspace{-1mm}
\subsection{Attack algorithms}
\vspace{-1mm}
\subsubsection{PGD and FGSM attacks}
\label{sec:pgd}
\vspace{-1mm}
The projected gradient descent (PGD) algorithm~\cite{madry2018towards} takes the benign audio waveform of length $T$, $\xvec\in\real^T$ 
and computes an adversarial example $\xvec'=\xvec+\deltavec$, by iteratively optimizing $\deltavec$ to maximize the misclassification error as 
\begin{align}
    \label{eq:iterfgsm}
   \deltavec_{i+1} = \mathcal{P}_\varepsilon(
   \deltavec_{i} + \alpha\,\mathrm{sign}(\nabla_{\xvec^{\prime}_{i}} L(g(\xvec^{\prime}_{i}), y^{\mathrm{benign}})) ) \;,
\end{align}
where function $g(\xvec)$ is the speaker/ASR classifier, $L$ is cross-entropy loss, $y^{\mathrm{benign}}$ is the true label, and $i$ is the 
iteration number. The function $\mathcal{P}_\varepsilon$ projects $\deltavec$ into the $l_p$ ball with radius $\varepsilon$. Usually, $p=\{1,2,\infty$\} are used 
with this attack. Typically, $\deltavec$
is initialized randomly, and the attacker tries several random initializations and uses the one providing the highest loss. 

There are simplified versions of this attack like the fast gradient sign method (FGSM)~\cite{Goodfellow2015} and 
Iterative FGSM (Iter-FGSM)~\cite{Kurakin2017}. Iter-FGSM is PGD-$L_\infty$ attack without
random initializations, i.e., $\deltavec$ is initialized to zero. FGSM is a a single iteration Iter-FGSM. 

\vspace{-1mm}
\subsubsection{Carlini-Wagner}
\label{sec:cw}
\vspace{-1mm}
The Carlini-Wagner (CW) attack~\cite{carlini-16} is computed by finding the minimum perturbation $\deltavec$ that fools the classifier while maintaining imperceptibility. $\deltavec$ is obtained by minimizing the loss,
   \vspace{-4mm}
\begin{align}
    \label{eq:cw}
    C(\deltavec) \triangleq D(\xvec,\xvec+\deltavec) + c \,f(\xvec+\deltavec)
\end{align}
where, $D$ is the distance metric of the threat model. $L_0$, $L_2$ or $L_\infty$ are used in the literature.
By minimizing $D$, we minimize the perceptibility of the perturbation. 
$f$ is defined in such a way that the system fails if and only if $f(\xvec+\deltavec)\le 0$. 
The precise definition for $f$ can be found in~\cite{Villalba2020a} and~\cite{joshi2021adversarial} for
attacks against speaker verification and classification respectively. 

\vspace{-2mm}
\section{Attack Representation Learning}

\subsection{x-Vectors}

In order to learn representations for attack classification/detection,
we propose to use the same kind of x-vector architectures that we use for speaker recognition.
The x-vector approach uses a neural network to encode the 
identity/attack information in each speech utterance into a single embedding vector~\cite{snyder-icassp18}.  
The x-vector network consists of 
three parts. First, an encoder network extracts 
frame-level representations from acoustic features (MFCC, filter-banks). 
This is followed by a global temporal pooling layer that produces 
a single vector per utterance--we used mean and standard deviation
pooling. Finally, a feed-forward network  computes attack class posteriors. 
The network is trained on a large set of attacks, using some form of cross-entropy loss. 
We employed additive angular margin softmax (AAM-softmax)~\cite{Deng2019} in this work. 
In the evaluation phase, the x-vector embedding is obtained from the first
affine transform after pooling, while the last layers of the network are discarded. Different x-vector
systems are characterized by different encoder architectures and pooling methods. In this work, we used a
Thin-ResNet34 encoder similar to the one in~\cite{Zeinali2019} with 16 to 128 channels in the 
residual blocks. 
In the context of adversarial attacks, 
we will denote the embeddings extracted from the x-vector network as \emph{attack signatures}.


\vspace{-1mm}
\subsection{Applications of attack signatures/embeddings}

\vspace{-1mm}
\subsubsection{Attack classification}
\vspace{-1mm}
This task entails classifying a test utterance into one of the known attack classes or the benign (unattacked) class. We can classify attacks attending to different criteria like optimization algorithm used to compute the adversarial perturbation, threat model, signal-to-noise ratio between the benign signal and the adversarial perturbation, etc. If the classes used to train the signature extractor network match our target classes, we can use the network output logits to classify the test sample. Otherwise, we can train another classifier, e.g., linear-Gaussian, PLDA, or logistic regression, on top of the signature vectors.

\vspace{-1mm}
\subsubsection{Attack verification/detection}
\vspace{-1mm}
Similar to speaker verification, attack verification is the task of deciding whether a test utterance contains the same attack as the enrollment utterance(s). In this case, we may have unknown attack types, i.e. attacks that are not included in the training of the signature extractor. We hypothesized that known attacks will define a manifold where unknown attacks also live. We used probabilistic linear discriminant analysis (PLDA)~\cite{Kenny2010} to evaluate the log-likelihood ratio between the \emph{same} vs \emph{different} attack hypothesis. 
\vspace{-1mm}
\subsubsection{Unknown attack detection}
\vspace{-1mm}
Unknown attack detection is the task of deciding whether a test utterance contains an attack not included in our training set. We used a PLDA model to compute the likelihood ratio between the \emph{unknown} vs \emph{known} attack hypothesis. We could prove that this is given by
 \vspace{-4mm}
\begin{align}
    \mathrm{LLR}&=
     -\log\frac{1}{N}\sum_{i=1}^N\frac{\Prob{\xvec_\mathrm{test},\Xmat_i|\mathrm{same}}}
    {\Prob{\xvec_\mathrm{test},\Xmat_i|\mathrm{diff}}}
\end{align}
where $\xvec_\mathrm{test}$ is the test signature; and $\Xmat_i$ are the signatures of the known attacks of class $i$.

\vspace{-2mm}
\section{Experiments}
\vspace{-1mm}
\subsection{Speaker and speech recognition tasks}

For speaker recognition, we created an experimental setup based on the VoxCeleb 1 and 2 datasets~\cite{Nagrani2020}.
The VoxCeleb2-dev set was split into VoxCeleb2-dev-train and VoxCeleb2-dev-test. For each 
speaker, we put 90\% of its utterances in dev-train and 10\% in dev-test.
VoxCeleb2-dev-train was used to train speaker recognition embedding networks. Meanwhile, 
VoxCeleb2-dev-test was used as the test set for a speaker classification task. The logits 
outputs of the speaker recognition network were used as scores for this task. 
We also used the standard VoxCeleb1-test-original-clean speaker verification task.
To evaluate speaker verification trials, we compared enrollment and test x-vector embeddings
using cosine scoring. Scores were calibrated into log-likelihood ratios by a logistic regression trained 
on benign trials.
x-Vector architecture was based on a Thin-ResNet34 similar to the ones used in~\cite{Zeinali2019,Villalba2020a}, with 256 embedding
dimension, additive angular margin=0.3.
As a reference, these systems provided 1.94\%, 1.91\%, and 3.2\% EER in the VoxCeleb1 original, entire and hard tasks.
We did not use a larger x-vector architecture to alleviate the huge computing cost of generating
adversarial attacks. In~\cite{Villalba2020a}, we showed that x-vector EER can degrade to 50\% (the worst possible) under
adversarial attacks, even with very high signal-to-perturbation-noise ratios ($>20$ dB). 

For automatic speech recognition (ASR), we trained our systems on the LibriSpeech~\cite{panayotov2015librispeech}
960 hours train split.  We tested on the first 100 utterances of the test-clean split, also to limit the computing cost. We evaluated Espresso~\cite{wang2019espresso} ASR system, based on the Transformer encoder-decoder architecture~\cite{vaswani2017attention}.
This setup is the same as~\cite{zelasko2021adversarial}.


\begin{table}
 \caption{Probability distributions used to generate attacks with random hyperparameters. In PGD-$L_1$, norms are
 normalized by the number of samples $n$, and in PGD-$L_2$ by $\sqrt{n}$, to make norm values comparable across utterances of different lengths}
    \label{tab:attack_hparam}
    \vspace{-3mm}
    \centering
    \resizebox{\columnwidth}{!}{
    \begin{tabular}{@{}lcc@{}}
    \toprule
    Algorithm & Hyper-parameter & Distribution \\ 
    \midrule
    (Iter-)FGSM/PGD-$L_x$     & max. $L_x$ ($\varepsilon$) & log-Uniform($3\times 10^{-6}$,$0.03$) \\
    Iter-FGSM/PGD-$L_x$     &  learn. rate ($\alpha$) & log-Uniform($10^{-6}$,  $0.005$) \\
    Iter-FGSM & num. iters.  & $1.25\; \varepsilon / \alpha$\\
    PGD-$L_x$ & num. iters. & Uniform(10, 100) \\
     & num. random inits & Uniform(2, 5) \\
    CW-$L_x$ & learn. rate & Uniform($10^{-5}$,$10^{-3}$) \\
             & confidence & Uniform(0,3) \\
    CW-$L_2$/$L_\infty$ & num. iters. & Uniform(10, 200) \\ 
    CW-$L_0$ & num. iters. & Uniform(10, 100) \\
         \bottomrule
    \end{tabular}}
    \vspace{-5mm}
\end{table}


\vspace{-1mm}
\subsection{Adversarial attack generation}

We generated attacks against VoxCeleb2-dev-train and VoxCeleb2-dev-test speaker classification; 
VoxCeleb1 speaker verification; and LibriSpeech-test ASR. For speaker recognition, we generated FGSM, Iter-FGSM, PGD-$L_\infty$/$L_1$/$L_2$ and CW-$L_\infty$/$L_0$/$L_2$ attacks using our own implementations.
To better model the variability of the attack manifold, the attack hyper-parameters were randomly sampled from different probability distributions as shown in Table~\ref{tab:attack_hparam}. For each attack algorithm-threat model,
we generated attacks for 25\% of utterances in the datasets. However, we only kept those attacks that were successful in
changing the classification label from correct to incorrect. For speaker verification, 
we used a decision threshold--$\theta=\logit\Ptar$ with target prior $\Ptar=0.05$, to classify trials as target or 
non-target. 

For ASR, we generated FGSM and PGD-$L_\infty$/$L_1$/$L_2$ attacks with norm values $\varepsilon=\{0.0001,0.001,0.01,0.1,0.2\}$. 
The experimental setup to generate ASR attacks is publicly available in the Armory toolkit\footnote{\tiny\url{https://github.com/twosixlabs/armory/blob/master/docs/scenarios.md}}, which uses the attack algorithms in the Adversarial Robustness Toolbox (ART)\footnote{\tiny\url{https://github.com/Trusted-AI/adversarial-robustness-toolbox}}.


\vspace{-3mm}
\subsection{Attack signature extraction networks}

The signature extraction networks were built with the same Thin-ResNet34 x-Vector architecture used for speaker recognition, 
but with embedding dimension empirically set to 10 and margin=0.2. We did not use noise and reverberation augmentation since it did not improve the attack classification performance.
The attacks and benign samples in VoxCeleb2-dev-train were used to train three signature extractors. Each of them was
discriminate a different attack property, namely,
attack algorithm+threat-model; threat-model; and signal-to-adversarial-noise ratio (SNR). Additionally, three extra networks were trained leaving out the CW attacks to evaluate the effect of having unknown attacks in the test set. 
These signature extractors, trained on speaker classification attacks, were evaluated on attacks against speaker classification and verification; and ASR.

\begin{figure*}
  \centering
  \begin{subfigure}{0.33\textwidth}
  \includegraphics[width=\textwidth,trim={12mm 8mm 16mm 14mm},clip]{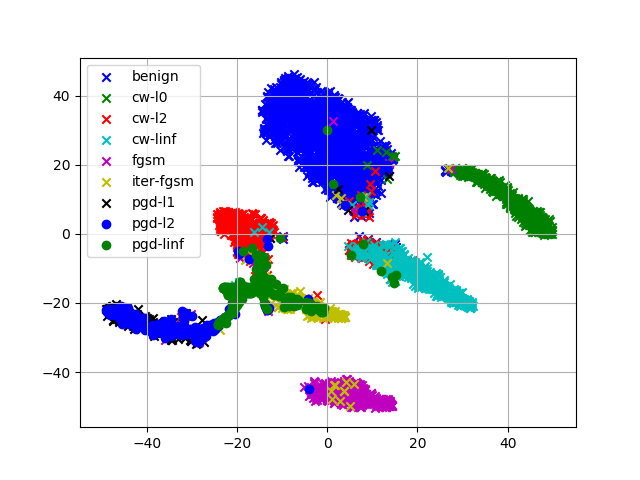}
  \caption{Attack algorithm}
  \label{fig:tsne_class_alg}
  \end{subfigure}
   \begin{subfigure}{0.33\textwidth}
  \includegraphics[width=\textwidth,trim={12mm 8mm 16mm 14mm},clip]{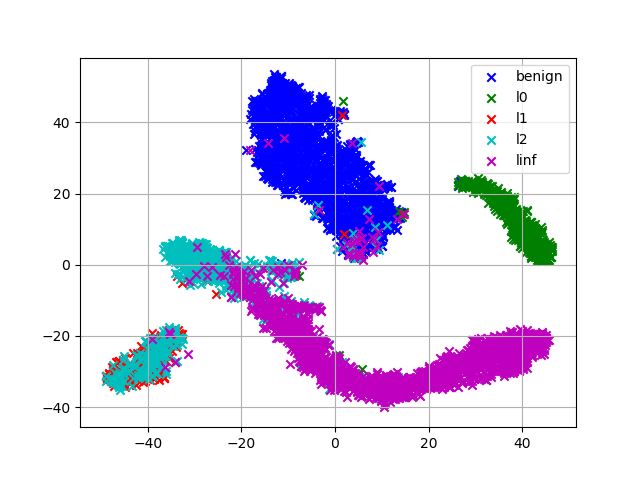}
  \caption{Threat model}
  \label{fig:tsne_class_tm}
  \end{subfigure}
   \begin{subfigure}{0.33\textwidth}
  \includegraphics[width=\textwidth,trim={12mm 8mm 16mm 14mm},clip]{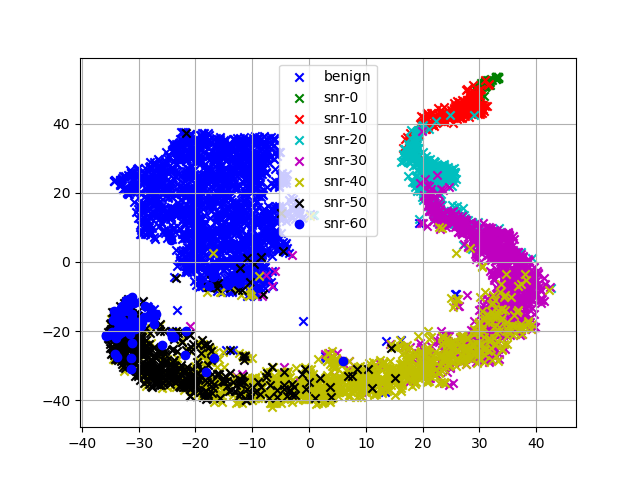}
  \caption{SNR(dB)}
  \label{fig:tsne_class_snr}
  \end{subfigure}
  \vspace{-4mm}
  \caption{T-SNE plots for attack signatures extracted from adversarial attacks on VoxCeleb2 dev-test speaker classification task. Signature extractors trained on our VoxCeleb2 dev-train set to discriminate attack algorithm, threat model or SNR.}
  \label{fig:tsne_class}
  \vspace{-5mm}
\end{figure*}

\setlength{\tabcolsep}{2pt}
\begin{table}
\caption{Normalized confusion matrix (\%) for attack algorithm classification in speaker recognition.}
    \label{tab:class_alg} 
    \vspace{-4mm}
    \resizebox{\columnwidth}{!}{
    \begin{tabular}{@{}c|ccccccccc@{}}
    \toprule
    &   Benign & CW-L0 & CW-L2 & CW-Linf & FGSM & Iter-FGSM & PGD-L1 & PGD-L2 & PGD-Linf  \\
    \midrule
    \multicolumn{10}{l}{\textit{VoxCeleb2-dev-test speaker classification attacks}} \\
    \midrule
    Benign & \bf 98.8 & 0.3 & 0.5 & 0.3 & 0.0 & 0.0 & 0.0 & 0.0 & 0.1  \\
    CW-L0 & 1.2 & \bf 98.6 & 0.1 & 0.2 & 0.0 & 0.0 & 0.0 & 0.0 & 0.0 \\
    CW-L2 & 2.8 & 0.1 & \bf 82.9 & 11.4 & 0.0 & 0.3 & 0.3 & 0.6 & 1.5  \\
    CW-Linf & 0.9 & 0.0 & 3.5 & \bf 95.0 & 0.0 & 0.0 & 0.0 & 0.0 & 0.6  \\
    FGSM & 1.5 & 0.1 & 1.1 & 0.1 & \bf 94.3 & 0.7 & 0.5 & 1.3 & 0.3 \\
    Iter-FGSM & 1.7 & 0.0 & 1.9 & 0.1 & 2.6 & \bf 82.4 & 0.0 & 0.3 & 10.9  \\
    PGD-L1 & 1.4 & 0.0 & 0.9 & 0.0 & 0.1 & 0.0 & \bf 91.4 & 6.2 & 0.0 \\
    PGD-L2 & 1.2 & 0.0 & 2.6 & 0.0 & 0.6 & 0.9 & \bf 46.5 & 42.7 & 5.6  \\
    PGD-Linf & 1.3 & 0.0 & 6.9 & 1.4 & 0.1 & 12.5 & 0.0 & 3.0 & \bf 74.6  \\
    \midrule
    \multicolumn{10}{l}{\textit{VoxCeleb1-test speaker verification attacks}} \\
    \midrule
    Benign & \bf 94.1 & 0.3 & 1.3 & 4.1 & 0.0 & 0.1 & 0.0 & 0.1 & 0.1 \\
    CW-L0 & 1.7 & \bf 97.9 & 0.1 & 0.3 & 0.0 & 0.0 & 0.0 & 0.0 & 0.0 \\
    CW-L2 & 4.5 & 0.2 & \bf 75.4 & 12.4 & 0.2 & 0.8 & 0.3 & 1.2 & 5.0 \\
    CW-Linf & 1.4 & 0.1 & 4.1 & \bf 93.5 & 0.0 & 0.0 & 0.0 & 0.0 & 0.9 \\
    FGSM & 3.0 & 0.1 & 2.7 & 0.2 & \bf 90.4 & 1.3 & 0.2 & 1.7 & 0.4 \\
    Iter-FGSM & 2.0 & 0.0 & 5.8 & 0.1 & 2.3 & \bf 74.9 & 0.0 & 1.6 & 13.3 \\
    PGD-L1 & 2.7 & 0.0 & 2.8 & 0.3 & 0.3 & 0.1 & \bf 85.2 & 8.6 & 0.1 \\
    PGD-L2 & 2.4 & 0.1 & 4.0 & 0.2 & 1.1 & 1.0 & 33.5 & \bf 49.8 & 7.8 \\
    PGD-Linf & 2.7 & 0.1 & 9.5 & 1.5 & 0.2 & 3.8 & 0.0 & 3.3 & \bf 79.0 \\
    
    \bottomrule
    \end{tabular}
    }
    \vspace{-6mm}
\end{table}

\setlength{\tabcolsep}{4pt}
\begin{table}
 \caption{Normalized confusion matrix (\%) for attack threat model classification in speaker recognition }
    \label{tab:class_tm} 
    \vspace{-3mm}

   \resizebox{\columnwidth}{!}{
    \begin{tabular}{@{}c|ccccc|ccccc@{}}
    \toprule
    & \multicolumn{5}{p{3.7cm}|}{VoxCeleb2-dev-test \newline speaker classification attacks} & 
    \multicolumn{5}{p{3.7cm}}{VoxCeleb1-test \newline speaker verification attacks} \\
    \cmidrule{2-11}
     & Benign & L0 & L1 & L2 & Linf & Benign & L0 & L1 & L2 & Linf\\
     \midrule
Benign & \bf 99.3 & 0.2 & 0.0 & 0.4 & 0.1 & \bf 96.1 & 0.2 & 0.0 & 1.4 & 2.3 \\
L0 & 1.3 & \bf 98.6 & 0.0 & 0.0 & 0.2 & 1.7 & \bf 98.1 & 0.0 & 0.1 & 0.1 \\
L1 & 1.4 & 0.0 & \bf 92.0 & 6.3 & 0.2 & 2.6 & 0.0 & \bf 88.0 & 9.0 & 0.3 \\
L2 & 3.1 & 0.0 & 17.6 & \bf 71.4 & 7.8 & 4.2 & 0.1 & 12.9 & \bf 71.2 & 11.6 \\
Linf & 1.7 & 0.0 & 0.2 & 9.6 & \bf 88.4 & 2.5 & 0.0 & 0.0 & 12.6 & \bf 84.8 \\
   \bottomrule
    \end{tabular}}
  \vspace{-4mm}
\end{table}

\begin{table}
\caption{Normalized confusion matrix (\%) for attack SNR(dB) classification in speaker recognition}
 \label{tab:class_snr} 
 \centering
 \vspace{-4mm}
 \resizebox{0.9\columnwidth}{!}{
 \begin{tabular}{@{}c|cccccccc@{}}
 \toprule
 SNR(dB) & Benign  &  0 &  10 &  20  & 30  & 40 &  50 &  60\\
 \midrule
 \multicolumn{9}{l}{\textit{Attack on VoxCeleb2-dev-test Speaker Classification Task}} \\
 \midrule
  Benign & \bf 98.7 & 0.0 & 0.0 & 0.0 & 0.2 & 0.5 & 0.5 & 0.3\\
    0 & 0.0 & \bf 87.3 & 12.7 & 0.0 & 0.0 & 0.0 & 0.0 & 0.0\\
   10 & 0.0 & 2.7 & \bf 90.8 & 6.4 & 0.0 & 0.1 & 0.0  & 0.0\\
   20 & 0.2 & 0.0 & 5.3 & \bf 84.3 & 10.1 & 0.2 & 0.0 & 0.0\\
   30 & 0.5 & 0.0 & 0.0 & 7.7 & \bf 78.8 & 12.9 & 0.0 & 0.0\\
   40 & 1.5 & 0.0 & 0.0 & 0.0 & 15.4 & \bf 66.9 &  16.1 & 0.1\\
   50 & 3.7 & 0.0 & 0.0 & 0.0 & 0.2 & 11.7 & \bf 80.7 & 3.7\\
   60 & 20.3 & 0.0 & 0.0 & 0.0 & 0.0 & 2.2 & \bf 50.6 & 26.9\\
 \midrule
 \multicolumn{9}{l}{\textit{Attack on VoxCeleb1-test Speaker Verification Task}} \\
 \midrule
 Benign & \bf 93.4 & 0.0 & 0.0 & 0.0 & 0.9 & 2.8 & 2.4 & 0.4 \\
 0 & 0.0 & \bf 79.4 & 20.6 & 0.0 & 0.0 & 0.0 & 0.0 & 0.0 \\
 10 & 0.0 & 1.3 & \bf  87.4 & 11.3 & 0.0 & 0.0 & 0.0 & 0.0 \\
 20 & 0.1 & 0.0 & 4.1 & \bf 78.5 & 17.1 & 0.2 & 0.0 & 0.0 \\
 30 & 0.3 & 0.0 & 0.0 & 5.4 & \bf 69.6 & 24.3 & 0.4 & 0.0 \\
 40 & 2.3 & 0.0 & 0.0 & 0.0 & 10.9 & \bf 64.7 & 21.8 & 0.2 \\
 50 & 5.0 & 0.0 & 0.0 & 0.0 & 0.4 & 11.9 & \bf 76.5 & 6.3 \\
 60 & 28.9 & 0.0 & 0.0 & 0.1 & 0.7 & 4.1 & \bf 45.0 & 21.1 \\
 \bottomrule
 \end{tabular}
 }
 \vspace{-5mm}
\end{table}

\vspace{-3mm}
\subsection{Attack classification in speaker recognition}

For these experiments, we used the three networks trained on all the attack algorithms. 
Table~\ref{tab:class_alg} shows confusion matrices\footnote{Confusion matrices have ground truth in rows, predictions in columns and are row normalized} for algorithm+threat-model classification.
For attacks against VoxCeleb2-dev-test speaker classification, we obtained 90.2\% attack
classification accuracy. Benign samples were correctly classified with 98.8\% Acc.; 
and most attack algorithms obtained more
than 74\% Acc. The largest confusion was between PGD-$L_2$ and $L_1$. 33\% of PGD-$L_2$ attacks were assigned to PGD-$L_1$. 
Thus, we found that is difficult to discriminate between $L_1$ and $L_2$ threat models. 
There was also significant confusion between PGD-$L_\infty$ and Iter-FGSM. However, this was expected since Iter-FGSM is just a
PGD-$L_\infty$ without random re-starts. The T-SNE plot in
Figure~\ref{fig:tsne_class_alg} presents similar evidence. We observe a significant overlap between PGD-$L_1$/$L_2$; and PGD-$L_\infty$ 
and Iter-FGSM; while the other classes are fairly well separated. 
For the attacks against VoxCeleb1 speaker verification, we obtained
84.6\% accuracy, and the confusion matrix follows a pattern similar to the previous one.
These results suggest that an attack classification model trained on a speaker classification task--closet-set of speakers, multi-class objective--, can be transferred to a verification task--open-set, unknown speakers, binary decision--, with small performance degradation.

Second, we look at the threat-model classification task. Here, accuracies were 90.7\% and 86\% for attacks 
against speaker classification and verification respectively. Confusion matrices in Table~\ref{tab:class_tm} show accuracies larger than 84\% for all threat models, except $L_2$, which is again 17\% misclassified with $L_1$. The T-SNE plot in Figure~\ref{fig:tsne_class_tm} shows us two clusters for $L_2$ threat model, one for CW-$L_2$ and another for PGD-$L_2$. The PGD-$L_2$ cluster is overlapped with PGD-$L_1$ cluster. The Benign, $L_0$ and $L_\infty$ clusters are well separated.
Again, we observe that the threat-model classifier transferred well from the speaker classification to the verification task.


Finally, Table~\ref{tab:class_snr} and Figure~\ref{fig:tsne_class_snr} present results for SNR classification.
Here, we note that most samples are classified in the correct SNR bin
or in the ones immediately next to it. The largest error was for attacks at SNR=60dB, which were 20-28\% misclassified as benign samples. However, few attacks were successful in fooling our speaker classifier at 60 dB (316), compared to 50 dB (1738) and 40 dB (3475).

\setlength{\tabcolsep}{2pt}
\begin{table}
\caption{Normalized confusion matrices (\%) for attack algorithm classification in ASR }
    \label{tab:class_alg_asr} 
    \vspace{-4mm}
    \resizebox{\columnwidth}{!}{
    \centering
    \begin{tabular}{@{}c|ccccc|cccc@{}}
    \toprule
    & \multicolumn{5}{c|}{With Benign} & \multicolumn{4}{c}{Without Benign}\\
    \cmidrule{2-10}
    & Benign & FGSM & PGD-L1 & PGD-L2 & PGD-Linf & FGSM & PGD-L1 & PGD-L2 & PGD-Linf \\
    \midrule
      Benign   &  31.2 & 1.0 & \bf 43.7 & 24.0 & 0.0 & - & - & - & -  \\
      FGSM     & 8.8 & \bf 71.8 & 1.5 & 14.3 & 3.6 & \bf 73.7 & 1.7 & 21.1 & 3.6\\
      PGD-L1  &   33.1 & 2.5 & \bf 46.3 & 18.2 & 0.0 & 3.7 & 47.1 & \bf 49.2 & 0.0\\
      PGD-L2 & 29.7 & 6.6 & 29.1 & \bf 30.1 & 4.5 & 8.5 & 29.7 & \bf 57.3 & 4.5 \\
      PGD-Linf & 11.1 & 3.2 & 1.1 & 21.9 & \bf 62.8 & 4.0 & 1.1 & 32.2 & \bf 62.8 \\
\bottomrule
    \end{tabular}}
\vspace{-4mm}    
\end{table}

\vspace{-2mm}
\subsection{Attack classification in speech recognition}

We evaluated our attack classifier on attacks against ASR, obtained 5\% accuracy
with about 50\% of attack files classified as benign. Then, we trained a PLDA classifier on top of 
the attack signatures using 100 ASR attack samples (20 per attack class), and tested on the remaining 1900 samples.
 When we consider benign samples in the experiment, accuracy was 50\%. 
 In Table~\ref{tab:class_alg_asr} (\emph{with benign}), we observe significant confusion between
benign, PGD-$L_1/L_2$, while FGSM and PGD-$L_\infty$ had better accuracies. If we do not consider
the benign class--assuming that we know that we are under attack--, accuracy raises to 60\%, but still
with large confusion between $L_1$ and $L_2$ threat models as we observed in the speaker recognition attacks.

We also evaluated our SNR classifier on the ASR attacks, obtaining promising results.
Here, 82\% of benign samples were correctly
classified and $>88$\% attacks with SNR$\in[0,40]$ were classified with error $\le 10$ dB. High SNR attacks were more problematic.
50 dB and 60 dB attacks were classified as benign with 30\% and 76\% probabilities respectively\footnote{\tiny we do not include full confusion matrix due to space constrains}. 

\begin{table}
    \centering
    \caption{EER(\%) for attack detection tasks. Attacks are against VoxCeleb2-dev-test speaker classification. Benign, FGSM and PGD are known attacks, CW attacks are unknown.}
    \label{tab:det}
    \vspace{-3mm}
     \resizebox{0.9\columnwidth}{!}{
    \begin{tabular}{@{}lccc@{}}
    \toprule
    & \multicolumn{3}{c}{Attack group} \\
    \cmidrule(l){2-4}
     Detection Task & Known+Unknown & Known & Unknown \\
     \midrule
     Algorithm+Threat-Model & \bf 19.6 & 12.6 & 41.5 \\
     Threat-Model & 24.8 & \bf 7.8 & \bf 36.6 \\
     SNR(dB) & 24.8 & 7.9 & 44.2 \\
     \bottomrule
    \end{tabular}}
    \vspace{-5mm}
\end{table}

\vspace{-2mm}
\subsection{Attack detection}

In this section, we approach the attack detection/verification task in a similar way as the speaker verification task. 
That is, given two utterances, we decide whether both contain the same or different attacks. We used a PLDA model
to compute the log-likelihood ratio between those two hypotheses.
For this experiment,
we used attack signatures and PLDA trained without CW attacks. Thus, benign, FGSM and PGD versions are known and CW versions are unknown. We created attack algorithm/threat-model/SNR verification trials using the adversarial samples obtained by attacking the VoxCeleb2-dev-test speaker classification task. In total, we obtained about 10M trials for each one of the attack verification tasks. 
Table~\ref{tab:det} shows the results in terms of equal error rate (EER). When considering only trials between known attacks, EER was low (8-12\%). However, for trials between unknown attacks, EER raised to 36-44\%--still better than chance. The best result was for threat-model detection with 36\% EER. This result indicates that signature extractors need to improve to be robust to unknown attacks. One possible direction is to train our signatures with a wider range of attack algorithms from those available in the literature. 


\vspace{-1mm}
\subsection{Unknown attack detection}

Here, we decide whether a test utterance contains an attack seen in training or not.  For this experiment, we used the signature extractor trained to discriminate attack algorithms on FGSM and PGD versions. CW versions were unknown attacks. A PLDA model was used to compute the likelihood ratio between the unknown and known attack hypothesis. 
We considered two conditions: with and without the benign class. In the former, the test utterance can be benign or attacked
and we
obtained 37.3\% EER. In the latter, all test utterances are adversarial--we assume that we have an oracle detector that tells us that we are under attack--, and 
we obtained 19\% EER. We think that this is a promising result that can be improved with further research.


\vspace{-2mm}
\section{Conclusions}

In this paper, we applied representation learning based on x-vector architectures 
to several tasks related to the classification and detection of adversarial attacks 
against speaker and speech recognition. First, we showed that the most common
adversarial attacks in the literature--FGSM, PGD, Carlini-Wagner-- can be detected and classified 
with accuracies as high as 90\%. We also showed that neural networks trained to classify attacks
against speaker identification can be used to classify attacks against speaker verification 
with small performance degradation. Furthermore, our learned representations termed as attack signatures
can be transferred to classify attacks against speech recognition. We also performed 
an attack verification task finding that attack signatures did generalize poorly 
to unknown attack algorithms. Finally, we evaluated an unknown attack detection task.
We found that unknown attacks can be detected with equal error rates of about 19\%, which is promising.


\bibliographystyle{IEEEtran}

\bibliography{mybib}


\end{document}